\documentclass[a4paper,onecolumn,11pt,accepted=2023-03-03]{quantumarticle}
\pdfoutput=1
\usepackage[utf8]{inputenc}
\usepackage[english]{babel}
\usepackage[T1]{fontenc}
\usepackage{amsmath}
\usepackage{hyperref}
\usepackage{doi}
\usepackage{dsfont}
\usepackage{tikz}
\usepackage{lipsum}
\usepackage{amsmath,amssymb}

\begin{document}

\title{Robust phase-controlled gates for scalable atomic quantum processors using optical standing waves}

\author{Shannon Whitlock}
\affiliation{European Center for Quantum Sciences and aQCess - Atom Quantum Computing as a Service, \\Institut de Science et d’Ingénierie Supramoléculaire (UMR 7006), University of Strasbourg and CNRS}
\orcid{0000-0002-6955-9326}
\email{whitlock@unistra.fr}
\maketitle

\begin{abstract}
A simple scheme is presented for realizing robust optically controlled quantum gates for scalable atomic quantum processors by driving the qubits with optical standing waves. Atoms localized close to the antinodes of the standing wave can realize phase-controlled quantum operations that are potentially more than an order of magnitude less sensitive to the local optical phase and atomic motion than corresponding travelling wave configurations. The scheme is compatible with robust optimal control techniques and spatial qubit addressing in atomic arrays to realize phase controlled operations without the need for tight focusing and precise positioning of the control lasers. This will be particularly beneficial for quantum gates involving Doppler sensitive optical frequency transitions and provides an all optical route to scaling up atomic quantum processors.
\end{abstract}

\section{Context and motivation}
Trapped atoms and atom-like systems are a leading candidate for numerous quantum technologies, including scalable quantum computing~\cite{Briegel2000, Saffman2016, Morgado2021}, optical atomic clocks~\cite{Ludlow2015} and quantum sensors~\cite{Pezze2018}. A key requirement in these applications is to reliably control the full quantum state of the system, typically by applying a sequence of phase-controlled optical pulses to the atoms. However, a major issue arises when scaling up to many atoms and large numbers of quantum operations, as errors associated to optical phase noise, atom position fluctuations and motion lead to rapidly accumulating errors~\cite{Saffman2011,Leseleuc2018, Graham2019, Day2022}.

A common way to realize phase-controlled quantum operations is by utilizing atomic transitions with frequency differences in the radio frequency or microwave regimes. These transitions can then be driven by stable microwave fields or a pair of Raman lasers with a small effective wavevector in a Doppler insensitive configuration~\cite{Wineland1998}. On the other hand, protocols involving optical frequency transitions~\cite{Norcia2019,Madjarov2019,Young2020,Schine2022} can have significant advantages when it comes to isolating the states and controlling differential light shifts, performing high fidelity state-resolved readout and mediating strong interactions exploiting highly-excited Rydberg states~\cite{Morgado2021}. However optical frequency transitions are generally much more sensitive to atom position fluctuations and motional dephasing making it difficult to realize high-fidelity phase-controlled operations.

Here a general method is proposed for realising robust and high fidelity quantum operations (gates) that is also compatible with spatial addressing of atomic qubits in large atomic arrays. By using two driving lasers in standing wave configuration it is possible to strongly suppress the sensitivity to local optical phase noise, including atom position variations and motion which are dominant errors in current experiments~\cite{Leseleuc2018,Picken2018,Graham2019}. Robust phase-compensated gates compatible with standing wave or travelling wave configurations can also be applied to individual qubits in large atomic arrays using targeted light-shifts~\cite{Labuhn2014,Wang2015,Wang2016}, where unwanted rotations on non-targeted qubits are cancelled by exploiting a time-reversal symmetry and geometric constraints. The proposed gates are robust, technically simple and parallelizable which will enable scaling up atomic quantum processors and high fidelity qubit control to large atomic registers.

\begin{figure}
    \centering
    \includegraphics[width=0.75\columnwidth]{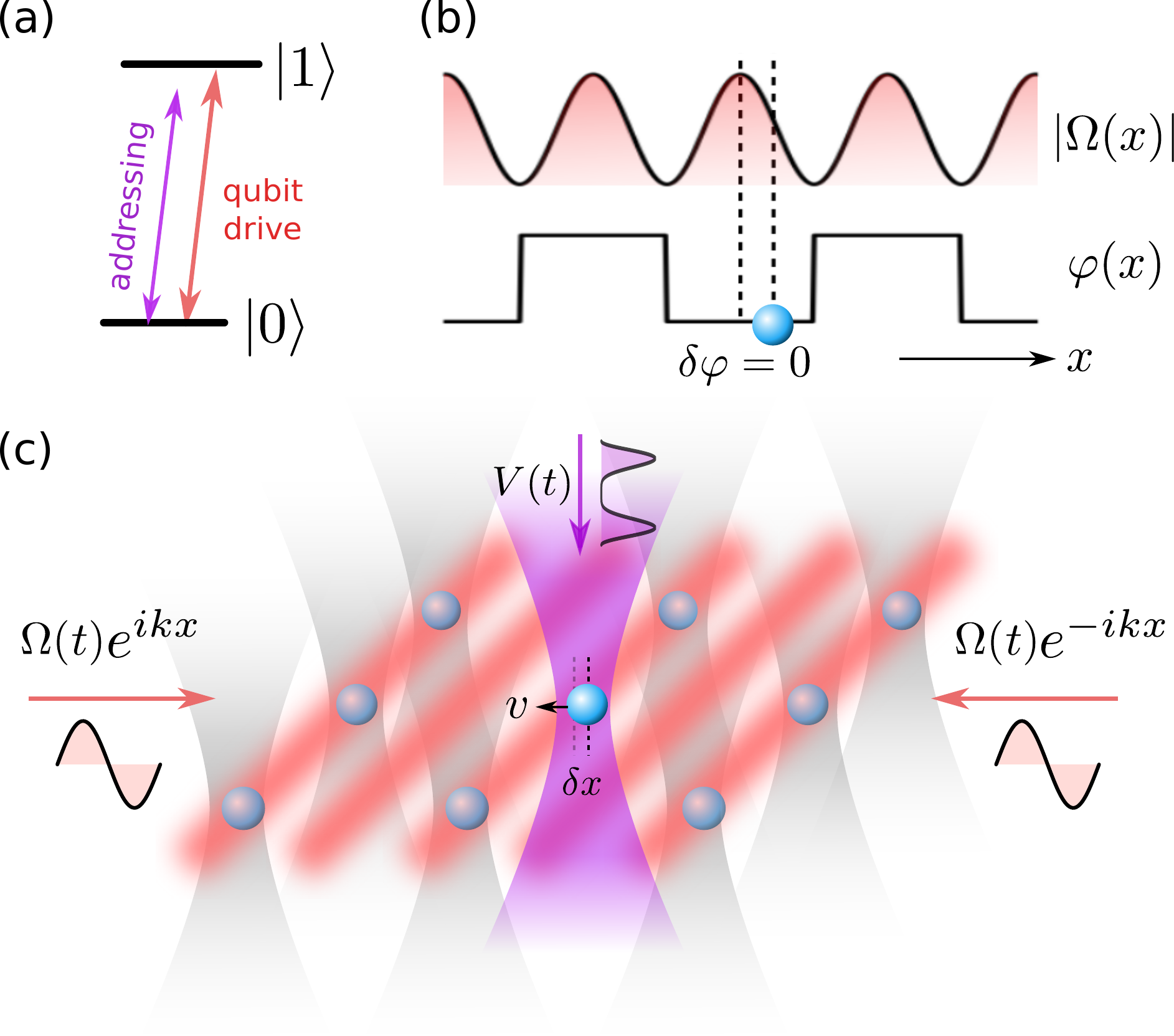}
    \caption{Setup for realizing robust phase controlled gates using optical standing waves. (a) Exemplary level diagram showing an optical transition for encoding an optical frequency qubit and laser fields for trapping and qubit addressing. (b) An interfering pair of counter propagating lasers produce an optical standing wave characterised by a periodically varying coupling strength $|\Omega(x)|$ and a phase $\varphi(x)$ which is spatially uniform between adjacent nodes. By controlling both the amplitude and phase of the lasers it is possible to realize phase-controlled quantum operations. (c) Scheme for single atom addressing involving a global standing wave drive field (red stripes) and an auxiliary far off-resonant beam (purple shaded region) targeting the central qubit.}
    \label{fig:fig1}
\end{figure}

\section{Phase-controlled gates using optical standing wave (OSW) fields}

Our goal is to realize phase-controlled qubit manipulations, which for a single qubit can be described by the following transformation acting on two atomic states
\begin{align}\label{eq:single_qubit_transformation}
    |0\rangle &\rightarrow \cos(\theta/2)|0\rangle -i\sin(\theta/2)e^{i\varphi}|1\rangle \nonumber\\
    |1\rangle &\rightarrow \cos(\theta/2)|1\rangle -i\sin(\theta/2)e^{-i\varphi}|0\rangle
\end{align}
Generally the rotation angle $\theta$ and qubit phase $\varphi$ is determined by the specific protocol used to drive the qubit. In the case of resonant driving $\theta$ is determined by the strength and duration of the atom-light coupling, while $\varphi$ is determined by, and very sensitive to the relative phase between the atomic dipole and the optical field at the position of the atom. 

To realize the transformation \eqref{eq:single_qubit_transformation} one can resonantly drive the atoms by two counter propagating optical fields with the same amplitude and phase forming optical standing wave (OSW). The drive Hamiltonian in the rotating wave approximation can be written
\begin{equation}\label{eq:hamiltonian_rwa}
    \hat H_d = \frac{\hbar}{2}\left(\Omega_1(t)e^{ikx}+\Omega_2(t)e^{-ikx}\right)|1\rangle\langle 0| + \textrm{h.c}
\end{equation}
where $\Omega_\alpha(t) = |\Omega_\alpha(t)|e^{i\varphi_\alpha(t)}$ are the complex valued Rabi frequencies of the two drive lasers with wavevectors $\pm k$, at the position of the atom $x$. For OSW gates $\Omega_1(t) = \Omega_2(t) = \Omega(t)$ is assumed.

This Hamiltonian allows to realize arbitrary phase controlled single qubit gates which are mostly insensitive to local optical phase noise and atomic position fluctuations. While an optical travelling wave (OTW) has a phase that is a linearly varying function of position $\propto kx$, an OSW on the other hand interferes to produce a field with a spatially uniform phase which steps between $\varphi$ and $\varphi+\pi$ between adjacent nodes (Fig.~\ref{fig:fig1}b). The atoms are assumed to be localized close to the antinodes of the OSW, either by making the atom array commensurate with the standing wave (Fig.~\ref{fig:fig1}c) or by shifting the relative phase of the drive fields between gates to address specific atoms. In this way it is possible to realize phase-controlled quantum operations that are mostly insensitive to the precise positions or velocities of the atoms. Instead the atom experiences a spatially varying intensity $\propto \cos^2(kx)$. But by tailoring the time-dependence of $|\Omega(t)|$ and $\varphi(t)$ (e.g., using electro-optic or acousto-optic modulators which can act on the timescale of several nanoseconds) it is possible to realize different quantum gates that also correct for associated intensity noise (or Rabi frequency) errors.

\begin{figure}
    \centering
    \includegraphics[width=0.75\columnwidth]{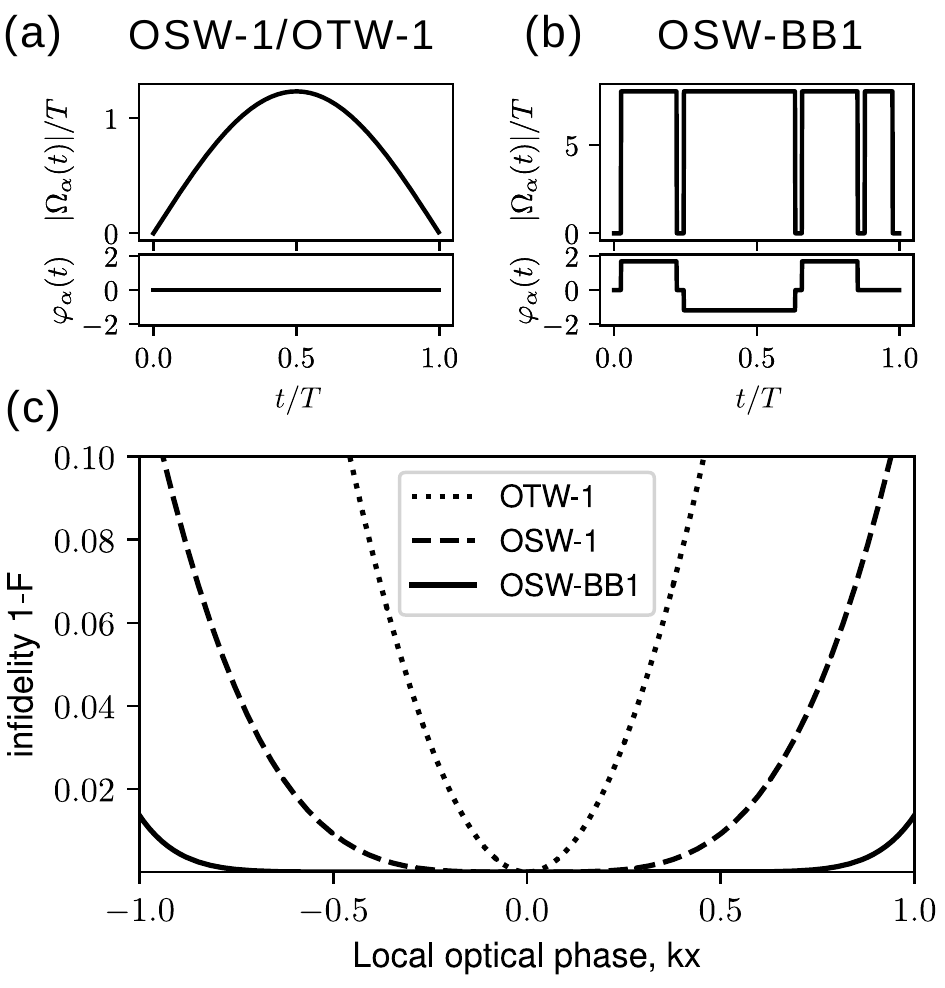}
    \caption{Infidelity of the $\text{R}_x(\pi/2)$ optical travelling wave gate (OTW-1) and optical standing wave (OSW-1 and OSW-BB1) gate sequences. (a) Amplitude and phase of the control field for a half-sine shaped control pulse with a combined pulse area of $\pi/2$ (for OTW-1 the control amplitude is scaled by a factor of 2). (b) BB1 pulse sequence which is optimized to cancel amplitude noise. (c) Comparison of gate infidelities as a function of the local optical phase $kx$ showing superior robustness of the OSW-1 and OSW-BB1 gate pulses.}
    \label{fig:fig2}
\end{figure}

To assess the advantage of OSW fields over OTW fields the achievable gate fidelity for three different gate protocols will be compared. The OTW-1 gate consists of a drive pulse $\Omega_1(t) =\nobreak A\sin(\pi t/T)$ and $\Omega_2(t) = 0$ with $A = \pi^2/(4T)$ (Fig.\,\ref{fig:fig2}a). The corresponding $\text{OSW-1}$ gate has $\Omega_1(t) = \Omega_2(t) = (A/2)\sin(\pi t/T)$ (note that the total required intensity for the OSW-1 gate is half that of OTW-1 for the same gate duration due to constructive interference). These simple parametric pulse shapes are favorable for experimental implementations, particularly for fast gates where bandwidth of the control systems might be limited. Finally the OTW-1 and OSW-1 gates are compared with a gate based on a four pulse BB1 sequence~\cite{Cummins2003} (OSW-BB1, see Fig.\,\ref{fig:fig2}b) which additionally suppresses sensitivity to Rabi frequency errors. In the absence of noise all three gates perfectly realize a $\textrm{R}_x(\theta = \pi/2)$ gate, equivalent to a $\sqrt{X}$ gate within a global phase factor, which is a basic operation for realising more complex multiqubit gates and quantum circuits. See Appendix~\ref{app:A} for generalizations to $U_{xy}(\theta,\phi)$ rotation gates according to Eq.~\eqref{eq:single_qubit_transformation}. Several $U_{xy}(\theta,\phi)$ gates can be concatenated to realize arbitrary single qubit control using OSW fields.

Figure~\ref{fig:fig2}c shows the gate infidelity $\epsilon = 1-F$ as a function of the local optical phase $kx$ assuming the atom is at rest during the gate time. $F$ is estimated by simulating the time-evolution operator $\hat U$ as a sequence of 400 piecewise constant segments and calculating $F=|\textrm{Tr}(\hat U_\textrm{target}^\dagger\hat U  )|^2/4$, where $\hat U_\textrm{target} = ((1,-i),(i, 1))/\sqrt{2}$. The simulations show the OTW-1 gate is most sensitive to the local optical phase with quadratic dependence on $kx$ with $\epsilon>0.1$ for $|kx| = 0.5$. In contrast the OSW-1 gate exhibits a less sensitive quartic dependence on $|kx|$ with $\epsilon=0.01$ for $|kx|=0.5$. The residual infidelity in the OSW configuration can be attributed to the spatial varying intensity, which can be corrected using more sophisticated pulse sequences. For example, the OSW-BB1 composite pulse sequence suppresses errors to 12th order in $|kx|$ with $\epsilon< 10^{-5}$ for $|kx| = 0.5$.

While the OSW configurations appear to be a vast improvement over the OTW configuration in the static case, it is important to know whether this improvement holds when the atoms move. To check this the Hamiltonian $\hat H_d$ is simulated with a time-dependent phase factor $kx(t)$ where $x(t)$ is obtained from the classical trajectory of an atom in a one-dimensional harmonic trap, i.e. $\dot x(t) = v(t), \dot v(t) = F(t)/m$ with $F(t) = m \omega^2 x(t)$, with random initial position $x(0)$ and velocity $v(0)$ sampled from a Maxwell-Boltzmann distribution, which is an approximation to the full quantum mechanical problem. Calculations are performed assuming realistic experimental parameters for $^{171}$Yb atoms, including a qubit transition wavelength of 578\,nm (standing wave period 289\,nm), trap frequency of $\omega/2\pi=100\,$kHz (trap period $\tau = 10\,\mu$s) and a temperature of $5\,\mu$K based on current experiments~\cite{Chen2022,Schine2022,Jenkins2022,Ma2022}. The average infidelity of each gate, for different gate times $T$, was obtained by averaging the infidelity $1-F$ obtained from the unitary evolution over 2000 classical trajectories. The Rabi frequencies of the OSW gates were scaled by a factor 1.03 to correct for the lower intensity seen by the atoms when averaging over their spatial distribution.

\begin{figure}
    \centering
    \includegraphics[width=0.65\columnwidth]{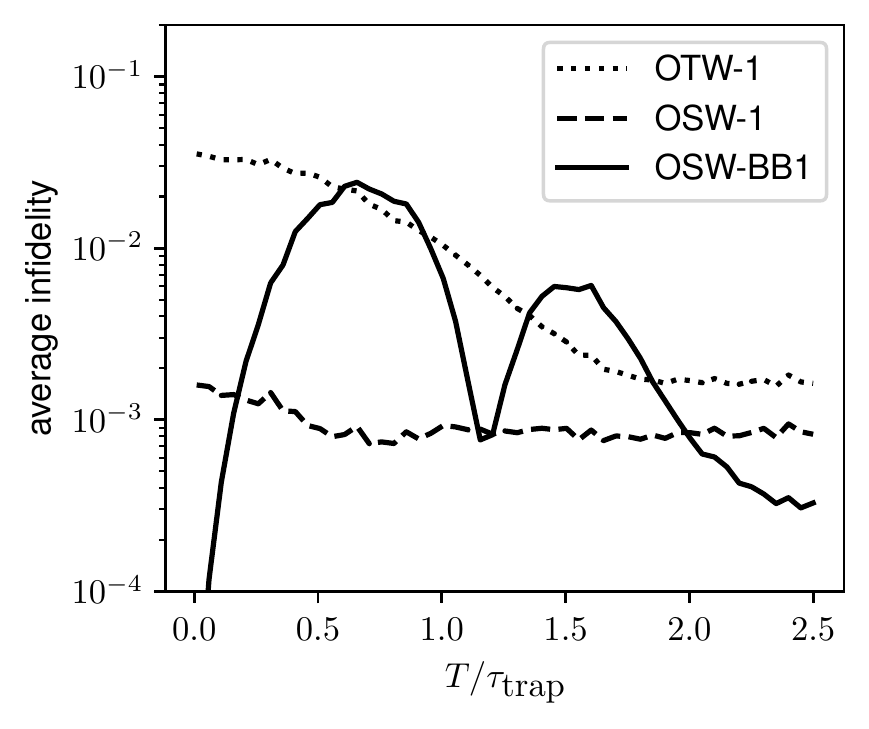}
    \caption{Average infidelity of $\text{R}_x(\pi/2)$ gates for a harmonically trapped $^{171}$Yb atom as a function of the gate time in units of the trap period. The position and velocity distribution is sampled from a thermal Boltzmann distribution with a temperature of $5\,\mu$K and trap frequency of $100\,$kHz which is then evolved according to the classical equations of motion. Each point corresponds to the mean of 2000 trajectories. The distribution of fidelities is highly non-Gaussian, with a median fidelity typically $3.7\pm 0.5$ and $6\pm 1$ times smaller than the mean and a standard deviation $3.0\pm 0.5$ and $2.6\pm 0.3$ times larger than the mean for the OSW-1 and OSW-BB1 gates respectively.}
    \label{fig:fig3}
\end{figure}

 Fig.~\ref{fig:fig3} shows the calculated gate infidelity for the OSW-1 gate is approximately 30 times smaller than the OTW-1 gate for $T\lesssim 0.5\tau$, and approximately a factor of 2 smaller for $T>1.8\tau$. The OSW-BB1 gate exhibits a higher sensitivity to atomic motion, however for very short and very long gate durations it can outperform the other two gates. Assuming a maximum Rabi frequency of $2\pi\times 200\,$kHz it is possible to obtain fast gates with $\epsilon_\textrm{OTW-1}\approx 0.032~ (T_\textrm{OTW-1}=2.0\mu$s), $\epsilon_\textrm{OSW-1}\approx 0.0014$ ($T_\textrm{OSW-1}=1.4\,\mu$s), and $\epsilon_\textrm{OSW-BB1}\approx 0.018$ ($T_\textrm{OSW-BB1}=8.8\,\mu$s). For the OSW gates the residual infidelity is dominated by trajectories with a large initial displacement which could be further improved using higher trap frequencies or lower temperatures.

\section{Robust qubit addressing using targeted light shifts}

Phase-controlled gates are also applicable to addressing in large atomic arrays without the need to tightly focus and spatially position the counter-propagating drive laser beams on individual atoms. The idea is to have all atoms interact with a common standing wave drive field while one or more target atoms are addressed by shifting their energy using an auxiliary focused and far-detuned laser~\cite{Labuhn2014,Wang2015, Wang2016}. This decouples coherent qubit control and spatial addressing and is applicable to 1D, 2D and 3D atomic arrays. Previous light shift addressing protocols for atomic qubits used focused lasers to induce targeted phase shifts~\cite{Labuhn2014, Wang2016} or to shift target qubits into resonance with a detuned drive field~\cite{Wang2015, Beterov2021}. However, in these examples one is typically limited to weak drive fields compared to the light shift or additional spin-echo sequences are required to correct unwanted rotations on the non-target atoms, resulting in slow gates. Here another approach is proposed for realizing fast and robust light-shift gates, which requires relatively small light shifts and guarantees the non-target qubits perfectly return to their original state at the end of the gate sequence, even in the presence of realistic noise sources. 
The basic idea is to exploit time reversal symmetry of the control Hamiltonian. For a resonant drive this can be achieved by making the drive field an anti-symmetric function of time,
\begin{equation}\label{eq:Omega_constraint}
    \Omega(t+T/2) = -\Omega(-t+T/2),
\end{equation}
which ensures that the non-target qubits return to their original state at $t=T$. 

For target atoms a local ac Stark shift of the form $V(t)|1\rangle\langle 1|$ is added to $\hat H_d$, where $V(t)$ is proportional to the addressing laser intensity and has a tailored time-dependence to realize a desired gate. For simplicity $V(t)\approx 0$ is assumed for non-target atoms. The same drive pulse $\Omega(t)$ can yield gates with different rotation angles $\theta$ enabling different gates to be applied to multiple qubits in parallel simply by shaping the addressing field. Multiple layers of gates with different phase angles $\varphi$ could be realized by concatenating pulses in the spirit of the sinusoidally modulated, always rotating, and tailored SMART qubit protocol recently demonstrated for spin qubits~\cite{Hansen2021a,Hansen2021b}, which can provide continuous noise protection.   

\begin{figure}
    \centering
    \includegraphics[width=0.8\columnwidth]{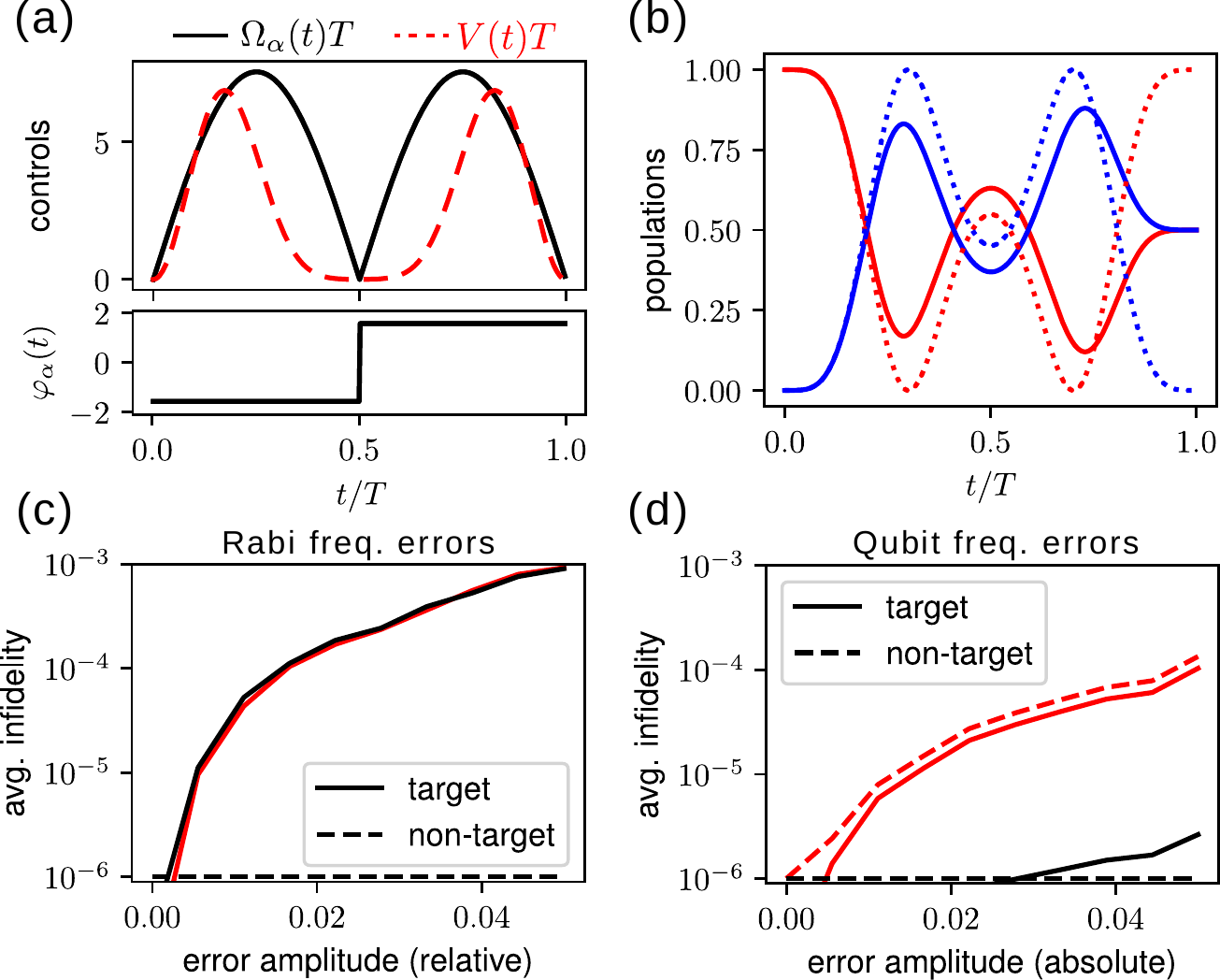}
    \caption{Robust gate protocol with light shift addressing. (a) Time-dependent controls for the OSW-LS2 gate. (b) Corresponding time-dependent populations for a target qubit (solid lines) and non-target qubit (dashed line) starting from the $|0\rangle$ state. (c) Average gate fidelity for target (solid black) and non-target qubits (dashed black) with respect to static Rabi frequency errors. (f) Similar comparison for static qubit frequency errors. The red lines show a comparison for a gate which is not corrected for qubit frequency errors (OSW-LS1, Appendix~\ref{app:A}).}
    \label{fig:fig4}
\end{figure}

Figure~\ref{fig:fig4}a,b presents a new gate OSW-LS2 suitable for addressing atomic qubits which has high robustness against both Rabi frequency and qubit frequency errors, especially for the non-target qubits. While the OSW configuration is analyzed, it should also be beneficial for travelling wave drives exploiting Doppler free resonances. The gate was numerically optimized using a restricted sum-of-sines basis with a cost function that provides second order cancellation of qubit frequency errors using the geometric formalism~\cite{Zeng2019} (further details in Appendix~\ref{app:B}). Besides its robustness to noise, a remarkable feature of the gate is that the required light shifts are comparable to, or smaller than the drive Rabi frequency which will be beneficial for realizing many gates in parallel. Fig.~\ref{fig:fig4}c,d shows the average infidelity (black lines) as a function of relative Rabi frequency errors (uncorrelated for each beam forming the OSW) and absolute qubit frequency errors (in units of $T^{-1}$) where averaging is performed over different static values of the Rabi frequency or qubit frequency sampled from normal distributions with standard deviation defined on the horizontal axis. Nearly perfect correction of both error types for non-target qubits are observed while target qubit errors are $<10^{-4}$ (Rabi errors) and $<10^{-6}$ (qubit frequency errors) for noise amplitudes as high as 2\% even though the gate is not directly optimized against target qubit errors. For comparison results are also shown for a non-corrected gate OSW-LS1 (red lines, details in Appendix~\ref{app:A}) which is significantly more sensitive to qubit frequency errors. For robustness with respect to intensity imbalance of the standing wave beams see Appendix~\ref{app:C}.

\section{Discussion}
In this paper several new gate protocols have been proposed for realizing high fidelity and spatially addressable quantum operations that are insensitive to qubit position fluctuations and motion, which is one of the major challenges affecting scalable quantum processors based on atomic qubits. This includes a new robust gate for addressing individual qubits that decouples control and addressing lasers, requires modest light shifts and perfectly corrects for errors on non-target qubits. Compared to travelling waves, standing wave drives can provide typically an order of magnitude lower infidelities approaching $10^{-3}$ for conditions achievable in current setups. The basic idea can be applied also to more complex multiqubit gates, including Rydberg mediated gates~\cite{Levine2019,Jandura2022}, which combined with phase controlled single qubit gates would form a universal gate set. It can also be applied to different geometries including standing-waves produced by optical cavities (where the spectral width of the gate pulses fit within the cavity bandwidth), holographic techniques or integrated photonic devices which may bring additional advantages including higher stability and faster quantum operations.

The OSW approach is especially advantageous for situations involving optical frequency transitions where traditional Doppler-insensitive excitation schemes like two-photon Raman transitions are not available. This brings forward the possibility of realising quantum computers based ultrastable optical clock qubits or new control strategies for optical atomic clocks and sensors that suffer less from the accumulation of errors associated with phase-control operations. 

\section*{Acknowledgements}
I thank S. Jandura, G. Pupillo, T. Bienaimé and Li Chang for valuable discussions. This work has benefited from a state grant managed by the French National Research Agency under the Investments of the Future Program with the reference ANR-21-ESRE-0032 and the Horizon Europe programme HORIZON-CL4-2021-DIGITAL-EMERGING-01-30 via the project “EuRyQa - European infrastructure for Rydberg Quantum Computing” grant agreement number 10107014 and support from the Institut Universitare de France (IUF). It is also part of the ITI 2021 2028 program of the University of Strasbourg, CNRS and Inserm, was supported by IdEx Unistra (ANR 10 IDEX 0002), and by SFRI STRAT’US project (ANR 20 SFRI 0012) and EUR QMAT ANR-17-EURE-0024 under the framework of the French Investments for the Future Program. Aspects of this work are described in disclosure "CNRS 16199-01" submitted as part of the ``Code Quantum'' maturation project supported by SATT Conectus. 
\bibliography{addressing}{}
\bibliographystyle{quantum}

\appendix
\section{Summary of gate protocols used in the paper}\label{app:A}

The gates reported in the main text can be used to realize general $U_{xy}(\theta, \phi)$ rotation gates of the form Eq.\eqref{eq:single_qubit_transformation} with the following control amplitudes.

\subsection{OTW}

\begin{align}
     \Omega_1(t) &= \frac{\theta\pi}{2T}\sin\left (\frac{\pi t}{T}\right )e^{i\varphi}, \\
     \Omega_2(t) &=0
\end{align}

\subsection{OSW}

\begin{align}
     \Omega_1(t) = \Omega_2(t) &= \frac{\theta\pi}{4T}\sin\left (\frac{\pi t}{T}\right )e^{i\varphi}, 
\end{align}

\subsection{OSW-BB1}

\begin{align}
\Omega_1(t) = \Omega_2(t) = \frac{\pi p}{2T}\times\left\{
        \begin{array}{ll}
            e^{i\phi_a} & \quad t(p/T) \leq 1 \\
            e^{i\phi_b}  & \quad 1 < t(p/T) \leq 3 \\
            e^{i\phi_a}  & \quad 3 < t(p/T) \leq  4\\
            e^{i\varphi} & \quad t/T \leq 1
        \end{array}
    \right.
\end{align}

with $p = \theta/\pi+4$, $\phi_a = \arccos(-\theta/4\pi)$ and $\phi_b = 3\arccos(-\theta/4\pi)$\cite{jones2009}.

\subsection{OSW-LS1 ($\theta = \pi/2$)}
\begin{align}
     \Omega_1(t) &= \Omega_2(t) = \frac{3.1317}{T}\sin\left (\frac{2\pi t}{T}\right )e^{i\varphi-i\pi/2}, \\
     V(t) &= \frac{3.5859}{T} \sin^2\left (\frac{\pi t}{T}\right )
\end{align}

\subsection{OSW-LS2 ($\theta = \pi/2$) with $\sigma_z$ cancellation}
\begin{align}
     \Omega_1(t) &= \Omega_2(t) = \frac{7.5551}{T}\sin\left (\frac{2\pi t}{T}\right )e^{i\varphi-i\pi/2}, \\
     V(t) &= \frac{1}{T} \left(2.1366\sin\left (\frac{2\pi t}{T}\right )+0.8875\sin\left (\frac{4\pi t}{T}\right )\right)^2
\end{align}

\section{Robust OSW-LS2 gate} \label{app:B}

The light-shift addressing concept can be applied also to more sophisticated OSW pulse sequences that can correct for different types of errors. For example, uncontrolled variations of the qubit frequency caused by residual light shifts from the tweezer traps or leakage of the addressing light on neighbouring qubits could be especially problematic as it potentially affects many qubits that interact with the drive field and not just the target qubits.

Finding a minimal basis and parameter sets which yield robust gates is a non-trivial problem, especially since the solution landscape can be very rugged. This problem was alleviated by restricting the optimization to a restricted sum-of-sines basis constrained to satisfy Eq.~\eqref{eq:Omega_constraint} and by adding an additional term to the cost function which favors solutions which are robust to perturbations.

Assuming the full Hamiltonian for non-targeted qubits can be written $\hat H(t) = \hat H_d(t) + \gamma \hat \sigma^z$, where $\sigma^z = |0\rangle\langle 0| - |1\rangle\langle 1|$ and $\gamma$ represents the qubit frequency difference with respect to the carrier frequency of the control field, which is treated as a small perturbation. The Hamiltonian in the interaction picture can be written as $\hat H_I(t) = \gamma \hat U_d^\dagger(t)\hat \sigma^z\hat U_d(t)$. The time-evolution induced by the perturbation can be determined from a Magnus series expansion $$\hat U_I(T) = \mathds{1}+\hat A_1(t) + \hat A_2(t) + \ldots$$ where $A_1$ and $A_2$ represent the first two orders of the expansion
\begin{align}
    \hat A_1(T) &= \frac{1}{\gamma}\int_0^T dt_1\hat H_I(t_1),\\
    \hat A_2(T) &= \frac{1}{2\gamma^2} \int_0^T dt_1 \int_0^{t_1} dt_2[\hat H_I(t_1),\hat H_I(t_2)],
\end{align}

For perfect noise cancellation, we desire $U_I(T)\rightarrow \mathds{1}$. In practice it is sufficient to cancel noise at first or second order, i.e. $\hat A_1(t) = 0$ or $\hat A_1(T) = \hat A_2(T) = 0$ respectively. In the geometric formalism~\cite{Zeng2019,Buterakos2021}, this is equivalent to the Bloch vector tracing out a closed curve (first order cancellation) with a projected area of zero in the xy, xz and yz planes (second order cancellation).

A simple control pulse shape that satisfies $\hat A_1(T) = \hat A_2(T) = 0$ and is subject to the time-reversal symmetry constraint is shown in Fig.\,\ref{fig:fig4}. It was found by minimizing $$\text{cost} = 1-F+||\hat A_1(T)||$$ using a restricted sum-of-sines basis for $\Omega(t)$ and $V(t)$, where $||\hat M||=|\text{Tr}(\hat M\cdot\vec \sigma)|$ with $\vec\sigma$ the Pauli vector $\{\sigma^x,\sigma^y,\sigma^z\}$. In practice, for the gates studied in this paper the symmetry constraint, together with $\text{cost}=0$ also ensures second order noise cancellation $\hat A_2(T) = 0$.

\section{Sensitivity to standing wave intensity imbalance} \label{app:C}
\begin{figure}
    \centering
    \includegraphics[width=0.8\columnwidth]{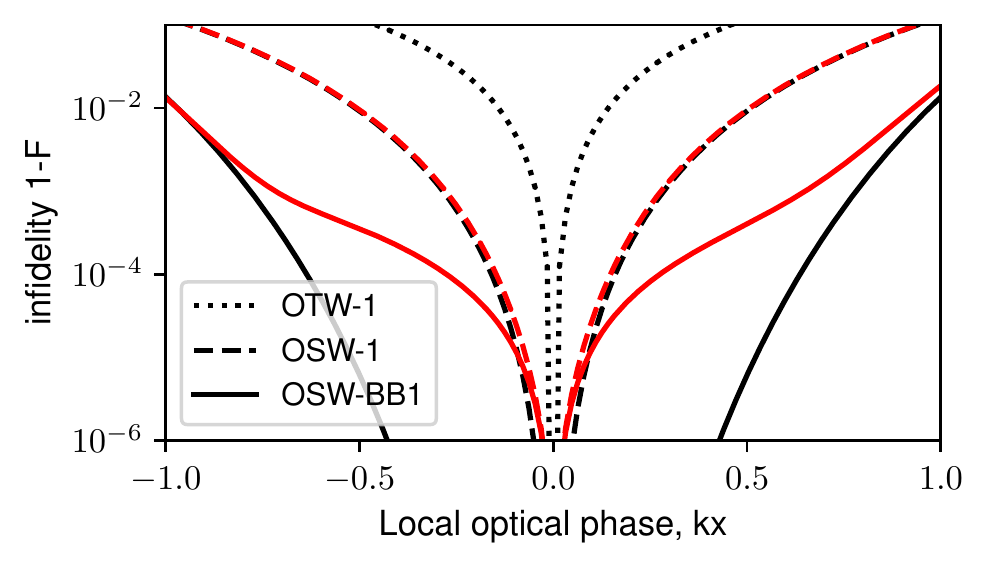}
    \caption{Infidelity of the $\text{R}_x(\pi/2)$ optical travelling wave gate (OTW-1) and optical standing wave (OSW-1 and OSW-BB1) gate sequences with (red lines) and without (black lines, same as Fig.~\ref{fig:fig2}c on a logarithmic scale) intensity imbalance, assuming a 20\% difference in intensities relative to the mean.}
    \label{fig:figS1}
\end{figure}

In practice it might be difficult to guarantee exactly equal intensities for the two beams which form the optical standing wave. This reduces the contrast of the standing wave and introduced a small gradient of the local optical phase at the antinode positions which can cause increased sensitivity to atom position fluctuations. To quantify this effect, Fig.~\ref{fig:figS1} shows the infidelity of OSW gates as a function of position (in units of the wavevector $k$), assuming a relatively large difference in intensities of 20\% relative to the mean. One sees that the effect on the OSW-1 gate is negligible, since the infidelity is dominated by the intensity inhomogeneity effect. In comparison, the OSW-BB1 gate which corrects for intensity homogeneity, is more sensitive to the intensity imbalance. Both OSW gates are an order of magnitude or more robust than the OTW gate and are compatible with infidelities $\sim 10^{-4}$ even in the presence of significant intensity imbalance.

\end{document}